\documentclass[preprint,aps]{revtex4}

\usepackage{graphicx}
\usepackage{dcolumn}
\usepackage{bm}
\newcommand{\bee}{\begin{eqnarray}}
\newcommand{\eend}{\end{eqnarray}}

\begin{document}

\preprint{APS/123-QED}

\title{Excitation and conversion of electromagnetic waves in pulsar
magnetospheres}

\author{G.~Gogoberidze$^{1,2}$, G.~Z.~Machabeli$^2$ and V.~V. Usov$^3$}
\affiliation{ $^1$ Centre for Plasma Astrophysics, K.U. Leuven,
Celestijnenlaan 200B, 3001 Leuven, Belgium \\$^2$Abastumani
Astrophysical Observatory, Ave. A. Kazbegi 2, Tbilisi 0160, Georgia
\\ $^3$Center for Astrophysics, Weizmann Institute, Rehovot 76100,
Israel \\}


\begin{abstract} We demonstrate that nonlinear decay of obliquely propagating
Langmuir waves into Langmuir and Alfv\'en waves $(L\rightarrow
L'+A)$ is possible in a one-dimensional, highly relativistic,
streaming, pair plasma. Such a plasma may be in the magnetospheres
of pulsars. It is shown that the characteristic frequency of
generated Alfv\'en waves is much less than the frequency of Langmuir
waves and may be consistent with the observational data on the radio
emission of pulsars.

\end{abstract}

\maketitle

\section{Introduction}
It is generally believed that a coherent (brightness temperatures up
to $\sim 10^{30}$~K) radio emission of pulsars is generated by
plasma instabilities in the one-dimensional, relativistic plasma
consisting of electron-positron pairs (e.g., \cite{M95}). Such pairs
may be produced by $\gamma$-rays via electromagnetic cascades in the
vicinity of the magnetic poles of pulsars that are identified with
strongly magnetized ($\sim 10^9-10^{13}$~G), rotation-powered
neutron stars. Created particles very rapidly lose the momentum
component transverse to the magnetic field and flow away along the
open magnetic field lines with Lorenz-factors ranging from
$\gamma_{\rm min}\sim 10$ to $\gamma_{\rm max}\sim 10^3$. A
two-stream instability has been proposed as a mechanism of radio
emission of pulsars soon after their discovery \cite{Ginzburg69}.
Later, using the available models of pulsar magnetospheres it was
argued that if the plasma outflow is stationary, the two-stream
instability does not have enough time to be developed before the
plasma escapes the pulsar magnetosphere (for a review, see
\cite{Usov02}). It was argued \cite{S71,RS75} that the process of
pair creation near the pulsar surface is strongly nonstationary, and
the pair plasma gathers into separate clouds spaced by $l\sim
10^6$~cm along the direction of its outflow. In this case, at the
distance $r_i\simeq 2l\gamma_{\rm min}^2\sim 10^8$~cm from the
neutron star fast ($\gamma\simeq \gamma_{\rm max}$) particles of one
plasma cloud overtake slow ($\gamma\simeq \gamma_{\rm min}$)
particles of the preceding cloud, and mutual overlapping of the
clouds begins. In the overlapping region, there are, in fact, two
steams of slow and fast particles, and the condition for the
development of two-stream instability is created \cite{Usov87}. For
typical parameters of pair plasma in the pulsar magnetospheres the
growth rate of the instability is quite sufficient for its
development (see below).

The frequency of Langmuir ($L$) waves generated at the distance
$r_i$ from the neutron star is a few ten times higher than the
typical frequency of pulsar radio emission (see \cite{Kunzl98} and
below). Therefore, the wave frequency has to decrease significantly
because of some processes to be compatible with the observational
data. In this paper, we discuss the process of conversion of $L$
waves into {\it low-frequency} Alfv\'en ($A$) waves that can solve
the high-frequency problem.

\section{Basic equations}\label{sec:2}

We consider pair plasma in the external uniform magnetic field ${\bf
B_0}$ directed along the $z$-axis. The behavior of this plasma may
be described by the Vlasov and Maxwell equations,
\begin{equation}
\frac{\partial}{\partial t} f_\alpha + v_z \frac{\partial}{\partial
z} f_\alpha +e_\alpha \left( E_z + \frac{1}{c} [{\bf v} \times {\bf
B}]_z \right) \frac{\partial}{\partial p_z} f_\alpha = 0,
\label{eq:01}
\end{equation}
\begin{equation}
{\bf \nabla} \times {\bf B} =\frac{4 \pi}{c}{\bf j} + \frac{1}{c}
\frac{\partial}{\partial t} {\bf E}\,,~~~~ {\bf \nabla} \times {\bf
E} - \frac{1}{c} \frac{\partial}{\partial t} {\bf B} = 0\,,
\label{eq:02}
\end{equation}
where $f_\alpha$ is the distribution function for the particles of
type $\alpha$ which is normalized so that $\int f_\alpha dp_z =
n_\alpha$, $n_0=\sum_\alpha n_\alpha$ is the particle density, ${\bf
B} \equiv {\bf B}_0+{\bf \tilde B}$, ${\bf \tilde B}$ is the
magnetic field perturbation, and $ {\bf j} = \sum_\alpha e_\alpha
\int {\bf v} f_\alpha d p_z $ is the current density.

Our study of nonlinear phenomena will be performed in the frame of
the weak turbulence theory. The particle distribution functions and
the current density may be written in the perturbation manner,
\begin{equation}
f_\alpha = f_\alpha^{(0)} + f_\alpha^{(1)} + f_\alpha^{(2)} + ...,
~~{\bf j} = {\bf j}^{(0)} + {\bf j}^{(1)} + {\bf j}^{(2)}+...,
\label{eq:05}
\end{equation}
here the unperturbed distribution function $f_\alpha^{(0)}$ is
assumed to be stationary and homogenous.

Besides, we use the infinite magnetic field approximation (for
finite magnetic field effects, see \cite{M80}). In this
approximation, the distribution function $f_\alpha$ is
one-dimensional in the velocity space (${\bf v}
\parallel {\bf B}_0$) and depends only on $v_z$. Technically, this
means that only the $z$ component of ${\bf j}^{(i)}$ is nonzero,
\begin{equation}
j^{(i)}_z = \sum_\alpha e_\alpha \int v_z f_\alpha^{(i)} d p_z\,.
\label{eq:07}
\end{equation}
In this case from Eqs. (\ref{eq:01}) and (\ref{eq:05}) we have
\begin{equation}
\frac{\partial}{\partial t} f_\alpha^{(0)} + e_\alpha \langle E_z
\frac{\partial}{\partial p_z} f_\alpha^{(1)} \rangle = 0\,,
\label{eq:08}
\end{equation}
\begin{equation}
\frac{\partial}{\partial t} f_\alpha^{(i)} + v_z
\frac{\partial}{\partial z} f_\alpha^{(i)} +e_\alpha\langle E_z
\frac{\partial}{\partial p_z} f_\alpha^{(i-1)}\rangle = 0\,,
\label{eq:09}
\end{equation}
where the angular brackets denote an average over one pulsation.

Performing Fourier transformation with respect to temporal and
spatial variables, Eq. (\ref{eq:09}) for $i=1,2$ yields
\begin{equation}
f_{\alpha}^{(1)}(\bar k) = -i e_\alpha \frac{E_{z} (\bar k)
\partial f_\alpha^{(0)}/\partial p_z }{\omega - k_z v_z},
\label{eq:10}
\end{equation}
\begin{equation}
f_{\alpha}^{(2)}(\bar k) = \frac{-e^2}{\omega - k_z v_z}
\frac{\partial}{\partial p_z} \int d {\bar k}^\prime E_{z}({\bar k}
- {\bar k}^\prime) E_{z}({\bar k}^\prime) \frac{\partial
f_\alpha^{(0)}/\partial p_z }{\omega^\prime - k_z^\prime v_z},
\label{eq:11}
\end{equation}
where $\bar k \equiv ({\bf k}, \omega)$. The first
($f_{\alpha}^{(1)}$) and second ($f_{\alpha}^{(2)}$) order
perturbations describe the linear dynamics of plasma and three wave
resonant processes, respectively.

Without lost of generality we assume that the wave vector ${\bf k}$
lies in the $x,z$-plane. From Eqs. (\ref{eq:02}), (\ref{eq:07}) and
(\ref{eq:10}) it can be shown that there are the following three
fundamental modes of pair plasma (e.g., \cite{VKM85}). The first one
is the extraordinary wave with the electric field perpendicular to
the ${\bf k},{\bf B}_0$-plane. The other two modes are $L$ waves
generated in the development of the two-steam instability and $A$
waves into which $L$ waves may be converted (see below). For both
these modes their electric fields lie in the ${\bf k},{\bf
B}_0$-plane.

In the infinite magnetic field approximation when only one component
($\varepsilon_{zz}$) of the linear permittivity tensor is nonzero
\cite{VKM85},
\begin{equation}
\varepsilon_{zz}=1- \sum_\alpha \omega_p^2 \int
\frac{F_\alpha}{\gamma^3(\omega-k_zv_z)^2} dp_z\,,\label{eq:13}
\end{equation}
for rather low wave numbers,
\begin{equation}
k_z^2 c^2 \ll 2\omega_p^2 \langle \gamma^{-3} \rangle,~~~ |k_z k_x|
c^2 \ll 2\omega_p^2 \langle \gamma^{-3} \rangle\,, \label{eq:17}
\end{equation}
the dispersion relations can be written in the form
\begin{equation}
\omega^2_L = 2\omega_p^2 \langle \gamma^{-3} \rangle +3 k_z^2 c^2
\left[ 1- \frac{\langle \gamma^{-3} \rangle}{\langle \gamma^{-5}
\rangle} \right] + k_x^2 c^2 \frac{k_z^2 c^2}{2\omega_p^2 \langle
\gamma^{-3} \rangle} \label{eq:15}
\end{equation}
for quasi-longitudinal $L$ waves, and
\begin{equation}
\omega^2_A = k_z^2 c^2 \left[1 - \frac{k_x^2 c^2}{2\omega_p^2
\langle \gamma (1+v_z/c)^2 \rangle} \right] \label{eq:16}
\end{equation}
for quasi-transversal $A$ waves, where
$F_\alpha=f_\alpha^{(0)}/n_\alpha$, so that $\int F_\alpha dp_z=1$,
$\omega_p =( 4\pi e^2 n_0/m_e)^{1/2}$ is the plasma frequency, and
$\langle ... \rangle \equiv \int ... F_\alpha dp_z$.

If the conditions (\ref{eq:17}) are fulfilled for $A$ waves the
ratio of their electric field components is
\begin{equation}
\Theta\equiv {E_z^A\over E_x^A}  \simeq \frac{k_xk_zc^2}{2\omega_p^2
\langle \gamma (1+v_z/c)^2 \rangle}\,. \label{eq:19}
\end{equation}

\section{Generation of $L$ waves and their evolution}\label{sec:3}

We consider the model where $L$ waves are generated by the
two-stream instability that develops at the distance $r_i\sim 10^8$
cm from the neutron star surface because of overlap of the pair
plasma clouds ejected from the pulsar \cite{UU88}. Below, all
calculations are done in the plasma frame where the mean
vector-velocity of the cloud particles is zero. We assume that in
the regions where the plasma clouds overlap the densities of the
slow and fast particles are the same (for development of two-stream
instability for arbitrary distribution function of plasma particles,
see \cite{UU88}). In this case the Lorentz-factor of the plasma
frame in the pulsar frame is $\gamma_p\simeq (\gamma_{\rm
max}\gamma_{\rm min})^{1/2}\simeq 10^2$.

The two-stream instability starts developing when the clouds
overlapping is very slight, and in the overlapping regions the
momentum spreads of the slow and fast particles are small
\cite{UU88}, i.e., the distribution function roughly is $F_\alpha
\simeq [\delta(p_z-p_0)+\delta(p_z+p_0)]$, where $p_0=\gamma_0 m_ec$
and $\gamma_0\simeq (1/2)(\gamma_{\rm max}/\gamma_{\rm
min})^{1/2}\simeq 5$. For $L$ waves with $k_x=0$ the dispersion
equation $\varepsilon_{zz}=0$ provides the maximum growth rate
$\Gamma_{max}\simeq \omega_p/(\sqrt{2} \gamma_0^{3/2})$ which is
achieved at $k_{z,opt}=\sqrt{6} \omega_p/(2 V_0 \gamma_0^{3/2})$,
where $V_0\simeq c$ is the particle velocity corresponding to the
momentum $p_0$. Obliquely propagating $L$ waves ($k_x\neq 0$) are
also generated by the two-stream instability. For such waves with
$k_x \lesssim k_z$ the growth rate is $\sim \Gamma_{max}$
\cite{M74}. The characteristic frequency of generated $L$ waves is
$\sim \omega_p \langle\gamma_0\rangle^{1/2}\simeq
\omega_p\gamma_0^{1/2}$ \cite{Usov87,UU88}.

When the amplitudes of generated $L$ waves become large enough
linear approximation is not valid any more, and the lowest order
nonlinear process called quasilinear relaxation starts up
\cite{LMS79}. This process may be described by Eq. (\ref{eq:08}),
and its characteristic time is
\begin{equation}
\tau_{QL} \simeq \Lambda \gamma_0^{3/2}/\omega_p\,, \label{eq:22}
\end{equation}
where $\Lambda$ is the Coulomb logarithm. The quasilinear relaxation
results in that the distribution function of particles has a shape
of a plateau up to the Lorentz-factor of $\sim \gamma_0$, i.e.,
$F_\alpha$ is $\sim \gamma_0^{-1}$ at $\gamma\lesssim \gamma_0$ and
nearly zero at $\gamma>\gamma_0$. The further generation of $A$
waves is cut off.

After the stage of quasilinear relaxation the induced scattering of
$L$ waves by plasma particles becomes the dominant nonlinear process
\cite{LMS79}. The characteristic time of this process is
\begin{equation}
\tau_{IS} \simeq \Lambda \frac{\gamma_0^{3/2}}{\omega_p} \frac{n_0
\gamma_0 m c^2}{W_L}, \label{eq:23}
\end{equation}
where $W_L$ is the energy density of $L$ waves. The induced
scattering transfers the wave energy from the frequency region where
$L$ waves are generated ($\omega \sim \omega_p \langle
\gamma\rangle^{1/2}\simeq\omega_p\gamma_0^{1/2}$ and $k_z\sim
k_{z,opt}$) to the low frequency region ($\omega \sim \omega_p
\langle \gamma^{-3} \rangle^{1/2}\simeq \omega_p\gamma_0^{-1/2}$ and
$k_z \sim \omega_p \sqrt{\langle \gamma \rangle}/c\simeq \omega_p
\gamma_0^{1/2}/c$), i.e., the mean frequency of $L$ waves decreases
$\sim \gamma_0\simeq 5$ times. Here, we used that $\langle
\gamma^{n} \rangle \sim \gamma_0^n$ for $n>0$ and $\langle
\gamma^{-n} \rangle \sim \gamma_0^{-1}$ for $n\geq 1$.

\section{Nonlinear decay  $L\rightarrow L'+A$}

The process of nonlinear conversion of $L$ waves into $A$ waves,
$L\rightarrow L'+A$ (and any other three-waves process) is described
by the second order current. From Eqs. (\ref{eq:07}) and
(\ref{eq:11}) this current can be written as
\begin{eqnarray}
j_z^{(2)} = \sum_\alpha \frac{e^3 \omega}{m} \frac{\partial
}{\partial p_z} \int d {\bar k}^\prime E_{z}({\bar k} - {\bar
k}^\prime) E_{z}({\bar k}^\prime)  \nonumber \\
\times\int dp_z \frac{\partial f_\alpha^{(0)}/\partial p_z
}{\omega^\prime - k_z^\prime v_z}
\frac{1}{\gamma^3(\omega-k_zv_z)^2}. \label{eq:24}
\end{eqnarray}
where $E_z=E_z^L+E_z^A$ in general case when both $L$ and $A$ waves
participate in the process.

The distribution functions of electrons and positrons created via
electromagnetic processes are identical. In such pair plasma all
three-waves processes are absent. This is because the second order
current ${\bf j}^{(2)}$ is proportional to $e^3$, and the summation
in (\ref{eq:24}) over electrons and positrons gives $j_z^{(2)}=0$.
However, in the process of outflow of pair plasma along the curved
magnetic field lines the distribution functions of electrons and
positrons are shifted with respect to each other to maintain the
electric neutrality of the plasma \cite{CR77}. This results in that
the mean Lorentz-factors of electrons ($\bar\gamma_e$) and positrons
($\bar\gamma_p$) slightly differ, $\Delta \gamma=|\bar\gamma_e
-\bar\gamma_p|\sim 1$.

For $L$ waves shifted to low frequency region by induced scattering
there is no solution of the resonant conditions,
$\omega_L=\omega_{L_1}+\omega_{L_2}$ and ${\bf k}_L={\bf
k}_{L_1}+{\bf k}_{L_2}$. Therefore, the process $L \rightarrow
L_1+L_2$ is kinematically forbidden, and we can omit all the terms
proportional to $E_z^L(\bar k) E_z^L({\bar k} - {\bar k}^\prime)$ in
Eq. (\ref{eq:24}). Taking into account that $\omega_L \gg k_Lc$ and
$\omega_A \approx k_A c$ for low-frequency $L$ waves and $A$ waves,
respectively,
from Eqs. (\ref{eq:19}) and (\ref{eq:24}) we obtain
\begin{equation}
-i(\omega-\omega_L) E_z^L = \int d {\bar k} V_{L^\prime A}
E_z^L({\bar k}-{\bar k}^\prime) E_x^A({\bar k}^\prime),
\label{eq:25}
\end{equation}
where
\begin{equation}
V_{L^\prime A} \approx \frac{4\pi e^3 n_0^2 \Delta\gamma \Theta}{m^2
c \omega_p \omega_A \gamma_0^{1/2}}, \label{eq:26}
\end{equation}
is the matrix element of interaction that  together with the
resonant conditions,
\begin{equation}
\omega_L=\omega_{L^\prime} + \omega_A~~~~{\rm and}~~~~ {\bf k}_L=
{\bf k}_{L^\prime}+ {\bf k}_A\,, \label{eq:27}
\end{equation}
totally determine the conversion process $L\rightarrow L'+A$.

\begin{figure}
\includegraphics[width=0.47\textwidth]{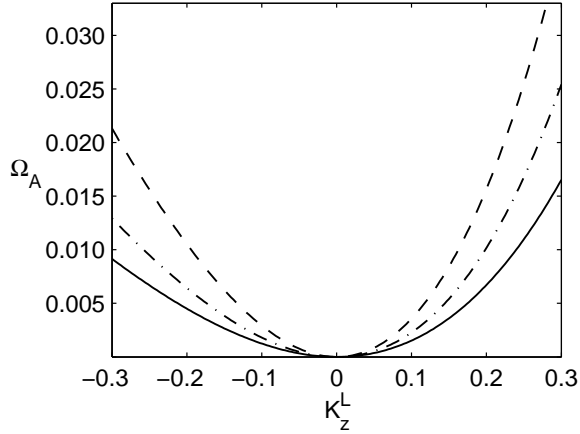}
\caption{\label{fig:1} Dimensionless frequency of $A$ waves
generated in the process $L\rightarrow L'+A$ as a function of
$z$-component of dimensionless wave number of $L$ waves before their
decay for different parameters: $K_x^L=0.5,~K_x^{L^\prime}=0$ (solid
line), $K_x^L=0.7,~K_x^{L^\prime}=0$ (dashed line) and
$K_x^L=0.7,~K_x^{L^\prime}=0.4$ (dash-dotted line).}
\end{figure}

To describe the frequency change in the process $L\rightarrow L'+A$
it is convenient to introduce the ration of the frequency of
generated $A$ waves to the mean frequency of the low-frequency $L$
waves, $\Omega_A \equiv \omega_A/(\sqrt{2}\langle \gamma^{-3}
\rangle^{1/2} \omega_p)\simeq
\omega_A/(\sqrt{2}\gamma_0^{-1/2}\omega_p)$. The results of
numerical solution of the resonant conditions (\ref{eq:27}) are
shown in Figure~1 where the dimensionless wave number is used,
$K^L_{x,y}\equiv
k^L_{x,y}c/(\sqrt{2\langle\gamma^{-3}\rangle}\omega_p)$. We can see
that for the typical parameters the frequency of generated $A$ waves
is about 2 orders of magnitude less then the mean frequency of $L$
waves. To clarify our numerical results presented in Figure~1 we can
find the approximate analytical solution of the resonant conditions
(\ref{eq:27}) for obliquely propagating $L$ waves, $\Omega_A \approx
(K_x^L)^2 (K_z^L)^2/2 \ll 1$ at $K_x^{L^\prime}=0$. From this
solution it follows that for rather small wave numbers of $L$ waves
the frequency of generated $A$ waves is always much less than the
characteristic frequency of $L$ waves.

Using the standard technique of weak turbulence theory (e.g.,
\cite{K65}), the characteristic time scale of nonlinear decay
$L\rightarrow L'+A$ for the waves with random phases may be written
as
\begin{equation}
\tau_{NLA} \sim \frac{\omega_L^2}{W_L V_{L^\prime A}^2 \omega_A}.
\label{eq:28}
\end{equation}

\section{Numerical estimates and discussion} \label{sec:4}

If we assume that in the plasma frame the frequency of radio
emission coincides with the mean frequency of $L$ waves generated by
the two-stream instability at the distance $r_i$, the characteristic
frequency of the radio emission in the pulsar frame is
\begin{equation}
\nu_L \sim 10\left( {B_{12}M \gamma_p \gamma_0/ P_{0.1}}
\right)^{1/2}\left( {R/r_i} \right)^{3/2} ~{\rm GHz}, \label{eq:29}
\end{equation}
where $B_{12}$ is the magnetic field at the neutron star surface in
units of $10^{12}~{\rm G}$, $P_{0.1}$ is the period of the pulsar
rotation in units of $0.1~{\rm s}$, $R\simeq 10^6$~cm is the neutron
star radius, and $M$ is a so-called multiplicity factor that equals
to the ratio of the pair plasma density to the density of the
primary particles \cite{Kunzl98}. In early papers where only the
curvature radiation of primary particles was considered as the
mechanism of generation of $\gamma$-rays the pair multiplicity is
rather high, $M\simeq 10^3-10^4$ \cite{RS75,DH82}. Later, it was
shown that generation of $\gamma$-rays by inverse Compton scattering
is essential near at least some pulsars \cite{HA01}. In this case
the value of $M$ may be as small as $\sim 1-10$. For plausible
parameters (e.g., $\gamma_p\sim 10^2$, $\gamma_0\sim5$, $M\sim
10-10^3$, $B_{12} \sim 1$, $P_{0.1} \sim 1$, and $r_i/R =50$) Eq.
(\ref{eq:29}) yields $\nu_L \sim 2-20~{\rm GHz}$ that is about an
order of magnitude larger then the typical frequency of the pulsar
radio emission.

In this paper we discussed the following sequence of processes in
the magnetospheres of pulsars: (a) development of two-stream
instability in collisions of plasma clouds and generation of $L$
waves; (b) quasilinear relaxation of two-stream instability; (c)
induced scattering of generated $L$ waves by plasma particles and
decrease of the mean frequency of $L$ waves; (d) nonlinear
conversion of $L$ waves into low frequency $A$ waves. For plausible
parameters of the pulsar plasma defined above and for
$K_{x,z}^L=0.3$ and $W_L\sim 0.1n_0 \gamma_0 m c^2$ the
characteristic times of these processes in the plasma frame are
$\tau_i=(\Gamma_{\rm max})^{-1}\sim 10^{-8}$~s, $\tau_{QL} \sim
10^{-7}$~s, $\tau_{IS} \sim 10^{-6}$~s, and $\tau_{NLA} \sim
10^{-6}$~s, respectively. These times are at least an order of
magnitude smaller than the characteristic time of the plasma outflow
$\tau_{\rm out}\simeq r_i/(c\gamma_p)\simeq 2\times 10^{-5}$~s.
Therefore, all these processes have enough time to be developed in
the magnetospheres of pulsars. We have shown that the frequency of
$A$ waves is a few ten times smaller than the frequency of generated
$L$ waves and may be consistent with the observational data on the
radio emission of pulsars.

Gedalin, Gruman, and Melrose \cite{GGM02} proposed another mechanism
of pulsar radio emission that also involves the two-stream
instability and solves successfully the high frequency problem at
least for the main part of known pulsars. In this mechanism
obliquely propagating electromagnetic waves are directly generated
by the instability. At present, the mechanism by Gedalin et al. has
an advantage over our mechanism because electromagnetic waves
generated by their mechanism can escape from the magnetospheres of
pulsars (provided cyclotron absorption is unimportant) while it was
argued (e.g., \cite{BA86}) that  $A$ waves are subject to strong
Landau damping and cannot be observed far from the pulsar (but see
\cite{B08}). However, in the simple model used in \cite{BA86} it was
assumed that the magnetosphere plasma is stationary, and the plasma
density declines in proportion to $r^{-3}$ without other variations.
These assumptions are invalid in our case where the pair plasma
outflow is nonstationary and strongly non-uniform across and along
the magnetic field lines \cite{RS75,Usov87}. At the edges of the
plasma clouds the density falls down sharply and significantly, and
$A$ waves may be converted into non-damping waves that escape as
radio emission. Besides, in the spaces between the plasma clouds the
density and Landau damping of $A$ waves may be small. A study of
these effects is beyond the framework of this paper and will be
addressed elsewhere.

In turn, our mechanism has an advantage over the mechanism by
Gedalin et al. for pulsars with the maximum of the radio spectra at
very low frequencies ($< 10^2$ MHz) because the mean frequency of
radio emission in our mechanism may be an order less than the same
in the mechanism by Gedalin et. al.

\begin{acknowledgments}
This work was supported in part by Georgian NSF grant 06-58-4-320,
INTAS grant 06-1000017-9258, and the Israel Science Foundation of
the Israel Academy of Sciences and Humanities.
\end{acknowledgments}


\begin{thebibliography}{20}
\bibitem{M95} F.C. Michel, {\it Theory of Neutron Star Magnetospheres}
(Univ. of Chicago Press, Chicago, 1991);  D.B. Melrose, J.
Astrophys. Astr. {\bf 16}, 137 (1995); M. Gedalin, D.B. Melrose, and
E. Gruman, Phys. Rev. E {\bf 57}, 3399 (1998).
\bibitem{Ginzburg69} V.L. Ginzburg, V.V. Zheleznyakov, and V.V.
Zaitsev, Astrophys. Space Sci. {\bf 4}, 464 (1969).
\bibitem{Usov02} V.V. Usov, in {\it Neutron Stars, Pulsars ans
Supernova Remnents}, edited by W. Becker, H. Lesch, and J. Tr\"mper
(MPE, Garching, 2002), pp 240-248; astro-ph/0204402.
\bibitem{S71} P.A. Sturrock, Astrophys. J. {\bf 164}, 529 (1971);
 Ya.I. Alber, Z.N. Krotova, and V. Eidman, Astrophizika {\bf
11}, 283 (1975); A. Levinson, D. Melrose, A. Judge, and Q. Luo.
Astrophys. J. {\bf 631}, 456 (2005).
\bibitem{RS75} M.A. Ruderman and P.G. Sutherland, Astrophys. J.
{\bf 196}, 51 (1975).
\bibitem{Usov87} V.V. Usov, Astrophys. J. {\bf 320}, 333 (1987).
\bibitem{Kunzl98} T. Kunzl, H. Lesch, A. Jessner, and A. Von
Hoensbroech, Astrophys. J. {\bf 505}, L139 (1998); D.B. Melrose and
M.E. Gedalin, Astrophys. J. {\bf 521}, 351 (1999).
\bibitem{M80} A.B. Mikhailovskii, Sov. J. Plasma Phys. {\bf 6}, 336
(1980); G.Z. Machabeli and G. Gogoberidze, Astron. Reports {\bf 49},
463 (2005).
\bibitem{VKM85} A.S. Volokitin, V.V. Krasnoselskikh, and G.Z.
Machabeli, Fiz. Plazmy {\bf 11}, 531 (1985).
\bibitem{UU88} V.N. Ursov and V.V. Usov, Astrophys. Space Sci. {\bf
140}, 325 (1988); E. Asseo and G.I. Melikidze, Mon. Not. R. Astron.
Soc. {\bf 301}, 59 (1998).
\bibitem{M74} A.B. Mikhailovskii, {\it Theory of Plasma Instabilities}
(Consultants Bureau, Plenum Publishing Corporation, New York, 1974),
Vol. 1.
\bibitem{LMS79} J.G. Lominadze, A.B. Mikhailovskii, and R.Z.
Sagdeev, Sov. Phys. JETP {\bf 77}, 1951 (1979).
\bibitem{CR77} A.F. Cheng and M.A. Ruderman, Astrophys. J. {\bf
212}, 800 (1977).
\bibitem{K65} B.B. Kadomtsev, {\it Plasma Turbulence} (Academic, New
York, 1965).
\bibitem{DH82} J.K. Daugherty and A.K. Harding, Astrophys. J. {\bf
252}, 337 (1982).
\bibitem{HA01} J.A. Hibschman and J. Arons, Astrophys. J. {\bf 560},
871 (2001); P.N. Arendt and J.A. Eilek, Astrophys. J. {\bf 581}, 451
(2002).
\bibitem{GGM02} M. Gedalin, E. Gruman, and D.B. Melrose, Phys. Rev.
Lett. {\bf 88}, 121101 (2002).
\bibitem{BA86} J.J. Barnard and J. Arons, Astrophys. J. {\bf 302},
138 (1986).
\bibitem{B08} A.M. Beloborodov, arXiv: 0710.0920.


\end{thebibliography}
\end{document}